# All-Optical Manipulation of Band Gap Dynamics via Electron-Phonon Coupling

Jicai Zhang[†], Tien-Dat Tran, Ziwen Wang, Wenhao Yu, Chong Zhang, Marcus Lo, Wenqi Xu & Tran Trung Luu[†].

[1]Department of Physics, The University of Hong Kong; Pok Fu Lam Rd, Hong Kong SAR, China.

[†]Corresponding author: jczhang@hku.hk; ttluu@hku.hk

**Abstract**

The electron-phonon coupling (EPC) is a ubiquitous interaction in condensed systems and plays a vital role in shaping the electronic properties of materials [1]. Yet, achieving coherent manipulation of electron-phonon coupling has posed a considerable challenge. Here, employing time-resolved high-harmonic generation (tr-HHG) spectroscopy, we demonstrate the coherent manipulation of bandgap dynamics in a *BaF$_2$* crystal by precisely controlling the EPC using ultrashort light pulses. The tr-HHG spectrum perturbed by a triply degenerate phonon mode T$_{2g}$, exhibits simultaneously a remarkable two-dimensional (2D) sensitivity, namely intensity domain in addition to the previously reported energy domain. The dynamic compression and enhancement of the harmonics in the intensity domain showed a $\pi/2$ phase shift compared to the manifestation of shifts of the harmonics in the energy domain, an astounding example of a physical phenomenon being observed simultaneously in two different perspectives. To complement our experimental observations, we employed a quantum model that incorporates the EPC, successfully reproducing the results. In addition, we demonstrated complete control over the EPC strength and initial phase of the coherent phonon oscillations by varying the incident electric field polarizations over crystal orientations. Our findings lay a foundation for future investigations aiming to harness and exploit the remarkable potential of EPC in solid-state systems.

Electrons, photons, and phonons are fundamental quantum components of condensed matter, with a comprehensive understanding and precise manipulation of electrons and photons serving as the cornerstone for technological advancements in modern society [1]. Significant progress has been made in developing high-speed electronic and optoelectronic devices [2-4]. However, to advance devices across various applications, it is crucial to comprehend not only the behavior of carriers but also their surrounding environment, specifically lattice vibrations known as phonons, and the ensuing interactions. The understanding, controlling, and practical applications of the phonons have remained challenging and elusive, leaving a significant gap in our current knowledge and capabilities [5-7]. Electron-phonon coupling, as a ubiquitous interaction in solid systems [8], plays a vital role in shaping the electronic properties of materials, such as electrical and thermal conductivities [9, 10]. The control of EPC holds significant potential for developing next-generation electronic devices characterized by improved performance and energy efficiency [5]. Furthermore, it offers a pathway towards exploring novel quantum phenomena and advancing functional [11] and superconducting [10] materials. The remarkable advancements in ultrafast laser technology [12] have facilitated unprecedented precision in investigating EPC within solid materials, offering a unique opportunity to explore the coherent manipulation of electronic properties. A myriad of time-resolved detection techniques has been developed to study dynamic processes [12] unfolding on the femtosecond timescale (1 fs = $10^{-15}$ s). Typically, coherent phonons can be generated by a pump pulse and probed using a separate pulse, enabling the detection of transient signal modulations [13, 14]. Previous investigations [15, 16] employing double-pump and single-probe transient spectroscopy have successfully harnessed control over the properties of coherent optical phonons, including amplitude and initial phase. However, the results from different coherent control experiments have relied upon disparate empirical theories [15, 16]. Consequently, there arises a need for a unified pump-probe experiment that integrates a microscopic quantum mechanical framework to elucidate the coherent control of EPC from diverse perspectives.

In recent decades, significant advancements in ultrashort light and electron sources have opened up new avenues for exploring the intricate dynamics of complex states of matter [17, 18]. High-harmonic generation (HHG), a well-established phenomenon in atoms and molecules, has emerged as a powerful tool for generating ultrashort sources in the extreme ultraviolet (EUV) to soft x-ray spectral range, thereby facilitating the study of matter dynamics on the attosecond timescale (1 as = $10^{-18}$ s) [19, 20]. High-harmonic spectroscopy [19] has proven successful in probing electronic structure and dynamics in atoms and molecules [21-24], as well as tracking chemical reactions in real-time [25, 26]. Notably, there have been recent breakthroughs, such as the work by Ghimire et al. [27], which extended HHG to solid-state systems, thereby enabling the generation of high harmonics from crystals like ZnO. This advancement in solid-state high-harmonic generation has unlocked remarkable possibilities for highly sensitive probing of material band structures [28], determination of crystal symmetries [27, 28], characterization of topological properties and material correlations [29-32], reconstruction of Berry curvature [33], and analysis of lattice vibrations [34-37], etc.

In this study, we employ time-resolved high-harmonic generation (tr-HHG) spectroscopy to trigger and monitor coherent phonon dynamics in *BaF₂* crystal. By manipulating the relative polarization and intensity of the pump light pulse, we demonstrate the ability to control the initial phase and strength of EPC. Our findings unveil that this precise control over EPC empowers us to modify the electronic band structure and dynamically modulate it through coherent phonons. Furthermore, we unravel the underlying mechanisms governing this observed coherent manipulation, providing invaluable insights into the fundamental physics of EPC in crystalline materials.

**Observation of coherent phonon oscillations in *BaF₂***

The concept of initiating and monitoring coherent lattice vibrations through tr-HHG has been predicted theoretically [35] and demonstrated experimentally [37]. We utilize a similar pump-probe configuration to investigate lattice properties of a dielectric x-cut Barium Fluoride (*BaF₂*) crystal (in [100] orientations), as depicted in Fig. 1a. The experiment involves the utilization of a high-intensity ultrashort pump pulse at a wavelength of 800 nm and a 400 nm probe pulse, arranged in a non-collinear pump-probe geometry. Time-delayed high-harmonic spectra are captured using an extreme ultraviolet (EUV) spectrometer. Detailed information regarding the experimental methodology can be found in the Methods section. The pulse duration of two pulses is measured to be around 22 fs (pump) and 25 fs (probe) with a homemade transient grating frequency-resolved optical gating (TG-FROG) device, as shown in the supplementary Fig. 1. Figures 1b and 1c display the supercell of the *BaF₂* crystal and its corresponding Brillouin zone (BZ). *BaF₂* is classified as a crystalline material belonging to the cubic crystal system. It possesses a face-centered cubic (fcc) lattice structure with a lattice parameter a of 6.23 Å. Within this lattice structure, the $F^-$ ions occupy tetrahedral sites, and each $Ba^{2+}$ ion is surrounded by eight $F^-$ ions, while each $F^-$ ion is surrounded by four $Ba^{2+}$ ions. Using density functional theory (DFT), we performed the calculation of the electronic structure and density of states of *BaF₂* crystal as shown in the supplementary Fig. 2. The obtained results reveal a direct band gap of approximately 10.2 eV at the center of the BZ, which closely agrees with the experimental measurement from neutron scattering (10.5 eV) [38]. Prior to introducing the pump pulse, we recorded the static high-harmonic generation (HHG) spectrum, as shown in Fig. 1d, which exhibits the 3$^{rd}$ and 5$^{th}$ harmonics (up to 15.5 eV). Notably, the former yields harmonics with intensities two orders of magnitude higher than the latter. By fixing the probe pulse at the P-polarized orientation and systematically rotating the crystal orientation, we successfully observed the cubic crystal symmetry structure. This is evident in Fig. 1e (H3) and 1f (H5), which reveal a periodic tetragonal crystal structure with a 90° periodicity. This symmetry corresponds to the symmetric path W-X-W within the BZ (see Fig. 1c).

The tr-HHG spectra we recorded are illustrated in supplementary Fig. 3a (H3) and 3b (H5). The traces exhibit a distinctive periodic beating pattern with a time-cycle of ∼ 140 fs (corresponding to a frequency of ∼ 7.2 THz), which directly arises from the presence of coherent lattice vibrations. To the best of our knowledge, this represents the first observation of coherent phonon dynamics in a dielectric *BaF₂* crystal. Notably, the tr-HHG spectrum demonstrates a two-dimensional (2D) sensitivity of the modulation characteristics. Signature of coherent phonons has been predicted theoretically as the modulation of harmonic spectral intensity [35, 36], and demonstrated experimentally as the modulation of harmonic photon energy [37]. In this work, both of these indications are measured simultaneously. This is illustrated by the dotted line in Fig. 2a and 2b, which represents the integrated harmonic yield spectrum and the center of mass (COM) of the 3$^{rd}$ harmonic (H3), respectively. The modulation encompasses both the compression and enhancement of harmonic yields (with a maximum modulated yield depth of ~4×10⁶) and the red and blue shifts of the harmonic central energy (with a maximum modulated energy depth of ~13 meV). To accurately capture and analyze the modulation signal, we fit the data using exponentially damped sinusoids. The solid blue line indicates in Fig. 2a and 2b that the modulation signal was fitted from exponentially damped sinusoids $S_i(t) = \Delta_i + A_i \sin[2\pi f_i t + \varphi_i] \exp[-t/\tau_i]$. Where the $\Delta_i$, $A_i$, $f_i$, $\varphi_i$, and $\tau_i$ indicate the zero offset, amplitude, frequency, initial phase, and relaxing time of the phonon (with "i" referring to the observed quantities), respectively. To further elucidate the distinctions between Fig. 2a and 2b, we normalize and plot them on the same time scale in Fig. 2c. As indicated by the two vertical dashed lines, a time delay of 36 fs is observed, which is equivalent to approximately one-quarter of the phonon oscillation frequency. To visually illustrate this intriguing observation, we employ a simple pendulum model in Fig. 2d. This model highlights the delay difference by introducing a π/2 initial phase discrepancy, where the maximum and minimum velocity (kinetic energy) of the pendulum correspond to the maximum modulation of harmonic yield and maximum modulation of harmonic energy, respectively.

The Fast Fourier Transform (FFT) of the two oscillation signals exhibits a similar pattern, as depicted in Fig. 2e, with a frequency of $f_i$ = 241.2 ± 0.5 cm⁻¹. This frequency value is in good agreement with our independent Raman measurement value of 241.1 ± 0.1 cm⁻¹, as indicated by the yellow line. Additionally, it is closely consistent with the calculated result of 238 cm⁻¹ obtained from density functional perturbation theory (DFPT) in supplementary Fig. 3a. Consequently, we confidently assign this peak to optical phonons with $T_{2g}$ symmetry, specifically a bending vibrational $T_{2g}$ phonon mode (supplementary Fig. 3b). This mode represents a triply degenerate state at the Γ point of the phonon dispersion. Typically, one would expect a similar phonon decay time across different relaxation channels. However, we observed an intriguing phenomenon in which the fitted relaxation time of the $T_{2g}$ mode differs between different observables. In particular, the phonon in the intensity domain exhibits a relaxation time of 833 ± 40 fs, which is longer than the relaxation time of 685 ± 30 fs observed in the frequency domain (not the Fourier-conjugate domain but the actual photon energy shift). This discrepancy may arise from the highly nonlinear nature of the HHG process in the detection spectrum, which involves multiphoton and electron tunneling processes within a short fs timescale.

**Quantum mechanical simulation**

To comprehend the temporal evolution of the physical processes underlying the tr-HHG radiation spectrum, we utilize numerical methods to solve both a two-level quantum model (see reference [37]) and a two-band model (see Methods part). The two-level model effectively captures the dynamics of the optically driven electron-hole pair system, which is coupled to coherent phonon oscillations.

$$\frac{d}{dt}f_0 = -i\varepsilon(t)(p - p^*), \tag{1}$$

$$\frac{d}{dt}p = -i\Omega(t)p + i\varepsilon(t)(1 - 2f_0) - \frac{p}{T_2}, \tag{2}$$

$$\Omega(t) = \frac{E_x}{\hbar} + \frac{V}{\hbar}\sin(\omega t + \varphi)\exp\left(-\frac{t}{\tau}\right) \tag{3}$$

with the exciton occupation $f_0$, the polarization $p$, the instantaneous Rabi frequency of the driving electric field $\varepsilon(t)$, the time-dependent exciton energy $\Omega(t)$ and the dephasing time $T_2$. $E_x$ represents the unperturbed exciting exciton energy, $V$ the *EPC* strength, $\omega$ the centre phonon oscillation frequencies, $\tau$ the phonon relaxing time, $\varphi$ the initial phonon phase. We examine the Fourier transform of the microscopic polarization $p$ as the source of the emitted radiation $E(\omega) \propto p(\omega)$.

To provide an intuitive description of the physical picture, as shown in Fig. 3a. When a pump pulse is introduced, the strong pump induces electronic transitions, causing a rapid modulation in the crystal potential experienced by the lattice atoms. This rapid shift triggers coherent phonon oscillations, which are subsequently detected through the coherent harmonic signal of the probe pulse [37]. By incorporating the appropriate fitted phonon parameters into our model, we simulated the dynamics of the coherent response, and the results are presented in Fig. 3c and 3d. The simulated spectra exhibit a remarkably close agreement with the experimental results. Therefore, through the simulation, we can gain an intuitive understanding of the physical processes involved by following these steps: (1) The excitation of carriers induced by the strong laser pulse leads to a distortion in the lattice potential, resulting in the generation of specific and coherent phonons. (2) These phonons then dynamically renormalize the band gap of the *BaF₂* through EPC process. (3) The modulation caused by the coherent phonons becomes evident in the harmonic yields and central energy of the tr-HHG signal of the probe. (4) The largest shift in the central energy of the HHG occurs when the exciton energy undergoes the fastest shift (at the zero position) and exhibits a $\pi/2$ phase delay difference in terms of HHG intensity yields. Note that the $\pi/2$ delay difference in the initial phase between the yields and central energy is accurately reproduced in our simulations. This implies that the anharmonicities of the $T_{2g}$ phonon mode, caused by phonon-phonon scattering, are very weak and can be neglected.

Usually, the generation mechanism of the coherent phonons can be classified into two main types: impulsive stimulated Raman scattering (ISRS) [15] and displacive excitation of coherent phonons (DECP) [39]. In conventional transient pump-probe experiments, it is common to only initiate phonons with A1 or E1 symmetry (doubly degenerate). In [39], the authors demonstrated that the observation of a pure symmetric A1 phonon gives the DECP generation mechanism. In our investigation, as the presence of a wide band gap (10.2 eV) in the *BaF₂* crystal, coupled with the involvement of multiphoton processes in HHG, strongly indicates that the DECP mechanism rather than ISRS, is primarily responsible for generating coherent phonons. This finding carries substantial significance as it underscores the generation mechanisms involved in DECP for the $T_{2g}$ mode. Furthermore, the observed 2D modulation in the tr-HHG spectrum not only provides a measure of the absolute strength of EPC but also offers a promising avenue for enhancing the yields of HHG through coherent phonon assistance. In both gas [19, 20] and solid [27, 28] media, maximizing the HHG yields is a key goal for optimal applications. Through various measurements and simulations, we have demonstrated that the phonon-assisted HHG process presents an exciting opportunity to achieve this objective. Through the utilization of an intensified light field, one can generate high-frequency phonon modes, thereby rendering it feasible to amplify the inherent modulation frequencies and strengths within the lattice structure. This, in turn, leads to an elevation in the efficiency of HHG yields across various time delays. Such an approach presents a promising pathway for augmenting the efficiency of HHG through the manipulation of EPC dynamics, similar to the concept of a "laser amplification" process.

**Manipulation of electron-phonon couplings via external light field**

To quantitatively assess the influence of EPC dynamics induced by light pulses, we fixed the intensity of the probe pulse at the order of $1\times10^{12}$ W/cm$^2$. The polarization of the probe pulse was aligned in the P-polarized direction along the X-W crystal orientation in the BZ. We recorded tr-HHG spectra while varying the polarization angle of the pump pulse. Fig. 4a illustrates the tr-HHG spectra obtained as we rotated the half-waveplate in the pump arm, ranging from 0° to 45° (corresponding to a polarization rotation from 0° to 90°). The dynamics of phonon oscillation are depicted by a diverse modulation in both amplitude and initial phase (or relative time delay) across the bottom to the top panel. The longitudinal optical (LO) and transverse optical (TO) splitting [40] arise from the interaction between light's electric field and charged particles in the crystal lattice, resulting in distinct energy levels for the LO and TO modes. The FFT spectra shown in Fig. 4b consistently exhibit a peak at 241.2 cm$^{-1}$, indicating that the polarization angles of the pulses do not significantly induce LO and TO mode splitting effects for the triply degenerate T$_{2g}$ phonon mode. Meanwhile, the EPC amplitude demonstrates noticeable modulations, as depicted in Fig. 4c. As the polarization angle transitions from 0° to 45° (corresponding to the high symmetric path from X-W to X-U in the BZ of *BaF$_2$*), the EPC strength varies from a maximum of 5.7 meV to a minimum of 0.9 meV before eventually fully recovering to its original strength around 5.6 meV. This behavior can be accurately fitted with a cosine curve, as indicated by the solid blue line in Fig. 4c. To gain a more intuitive understanding of the underlying physics, we conducted DFT calculations and generated a 2D electron density distribution on the *Ba* atoms within the x-cut (i.e., [100]) plane, as shown in Fig. 4d. The calculations specifically capture the density of valence electrons occupying orbitals below the Fermi level (E = 0 eV). The T$_{2g}$ mode, whose generation mechanism is attributed to the DECP, exhibits a profound link between the amplitude of phonons and the electron density $n_e(t)$. In the classical trajectory analysis which elucidates electron excitations in physical space, the connection between electrons and the illuminating electromagnetic field is established via the orientation-dependent transition dipole moment *d(θ)*. This moment serves as a measure of the probability amplitude for an electron to traverse between two quantum states under the influence of the electromagnetic field. The direct correspondence between the interaction of electrons with the light field and the transition dipole moment *d(θ)* indicates that the strength of this interaction hinges upon the alignment and orientation of the dipole moment to the incident light field. This connection leads to the dynamic tetragonal variation of EPC strength in response to changes in pump polarization concerning different crystal orientations.

In addition to the observed variations in EPC strength, we have also discovered dynamic changes in the initial phase corresponding to the generation time following the illumination of the pump pulse. When considering polarization angles ranging from 0° to 45°, there is a gradual decrease in the relative time delay, spanning from 0 to 5 fs, which aligns with a transition from 0 to -0.3π. However, beyond the 45° mark, the time delay experiences a rapid increase, eventually reaching 60 fs. This sharp flip signifies an almost π shift in the initiation phase of phonon generation. Subsequently, as the polarization angle continues from 45° to 90°, the time delay exhibits a gradual decrease once again. This phenomenon can also be lucidly explained by considering the electron's trajectories within different crystal orientations. As the electron undergoes diverse temporal excursions, the durations of these excursions vary in different orientations, leading to distinct initiation times (or phases) for the perturbation of the lattice.

Within the energy domain, the rotation of polarization introduces variations in exciton energy due to the K-dependent energy dispersion. Our two-level model establishes a correspondence between different polarization angles and distinct exciton energies. Supplementary Fig. 4 provides a visual representation of this relationship. Through simulations involving the manipulation of the exciting exciton energy, we have successfully achieved precise control over the shape of the spectrum, which characterizes the initial phase. By finely tuning the exciting exciton energy within the range of 9.0 eV to 9.5 eV, we have effectively achieved an almost 2π phase modulation. Of particular interest is the behavior observed when the exciting energy of the exciton ($E_x$) approaches the energy of the probe's THG at 9.3 eV. In such cases, the resulting spectral trace exhibits a distinct asymmetrical shape. Conversely, when the energy deviates significantly from both sides of the THG energy, the tr-HHG signal takes on a more symmetric profile. Additionally, our results encompass the dynamics of EPC, wherein we recorded tr-HHG spectra under varying pump intensities (~$1\times10^{12}$ to $5\times10^{12}$ mW/cm²) and probe intensities (~$1\times10^{12}$ to $2\times10^{12}$ mW/cm²) for the pump pulse. Supplementary Fig. 5a and 5b show the tr-THG spectra as the intensity of both the pump and probe is increased while maintaining a fixed polarization of P-polarized. The modulation depth of both harmonic yields and central energy exhibits a linear relationship with increasing pump fluence or decreasing probe fluence.

In comparison to the conventional double-pump and single-probe technique [15, 16], the tr-HHG-based approach offers the notable advantage of employing a single-pump and single-probe configuration. Coupled with the inherent 2D sensitivity in capturing phonon responses, our results establish a solid foundation for the coherent manipulation of electron-phonon coupling (EPC) dynamics, both in the initial phase and amplitude. The combination of experimental and simulated results has enabled us to demonstrate the coherent manipulation of the initial energy position within the electronic structure of *BaF$_2$* on a timescale fs. This manipulation is achieved by precisely controlling the EPC through the utilization of different polarization angles for the pump pulses. The ability to modulate EPC through pump polarization provides a means to exert control and engineer the band gap by interactions between electrons and lattice vibrations.

**Discussion**

To investigate a coherent perturbation induced by lattice vibrations, we move beyond a simple two-level model and adopt a more comprehensive two-band model to unravel the fundamental mechanisms driving HHG in solid materials. By employing tr-HHG spectroscopy, we can delve into the interplay between electronic bands and lattice vibrations, providing insights into the underlying processes that govern HHG in solids. Current theories, based on a single-electron approximation, propose that the primary mechanisms contributing to HHG in solids are interband polarization and intraband current. However, there is an ongoing debate regarding the dominant mechanism, whether it is the intraband current $J(\omega) \propto \partial \Delta \varepsilon_k / \partial k$ (related to the energy bands) [27], interband polarization $p(\omega) \propto d(k',k)$ (related to the transition dipole moment) [28], or a combination of both [30]. These mechanisms assume the absence of correlations between electrons and other particles, such as electrons, holes, and phonons. Nevertheless, it is crucial to consider the presence of correlation effects inherent in complex solid systems, including electron-electron interactions and electron-phonon interactions. By incorporating these correlation effects, particularly electron-phonon correlations, into a two-band system, we find the ability to differentiate between interband and intraband HHG components. In our analysis, we take into account the coherent phonons, generated with a wave vector **Q** = 0, and consider the phonon-related terms in the equation of motion based on the semiconductor Bloch equations (SBEs). For a more detailed description of the phonon-perturbed SBEs, see quantum two-band model section.

We incorporated the $T_{2g}$ phonon mode of *BaF$_2$*, which was utilized in the simulation, into our phonon-perturbed two-band model. To explore a wider range of harmonic orders, we adjusted the center wavelength of the probe pulse to 800 nm. The corresponding results are presented in supplementary Fig. 6a, where we observe that all tr-HHG signals exhibit the same initial phase. However, each harmonic order (from the 5$^{th}$ to the 15$^{th}$) displays a distinct response in terms of modulating amplitude (supplementary Fig. 6b and Fig. 6c). This behavior is expected, as high-harmonic generation from solid materials is intricately connected to the underlying band structure of the material. Interestingly, we observed a notable difference between the contributions of interband polarization and intraband current to the tr-HHG spectrum response. As depicted in Fig. 5a and 5b, the interband polarization exhibits a greater modulation strength compared to the intraband current in the harmonic energy domain, while maintaining the same initial phase. This disparity becomes particularly pronounced when the bandgap is adjusted to be close to the energy of THG. Leveraging the deformation potential theory of the EPC, we propose a potential methodology for discriminating between the main contributions of interband and intraband harmonics. This method involves measuring the tr-HHG spectrum in a system perturbed by coherent phonons. Specifically, we suggest selecting a probe field with one of the harmonic orders that closely aligns with the bandgap of the given material.

**Conclusion**

In conclusion, our study utilizes the 2D tr-HHG spectroscopy technique to explore the dynamics of the triply degenerate $T_{2g}$ phonon mode in the dielectric *BaF$_2$* crystal, shedding light on the behavior of the EPC. This 2D sensitivity encompasses the time-dependent compression and enhancement of harmonic yields, as well as the intriguing red and blue shifts observed in the central energy of the harmonics. By integrating the EPC and first-principles DFT calculations into our quantum model, we successfully reproduce the observed 2D modulated tr-HHG spectra. Moreover, our findings unveil the remarkable capability of precise manipulation of band gap dynamics through a coherent external light field, providing effective control over the strength and excitation phase of the EPC in $BaF_2$. This newfound control over the EPC enables coherent modifications of the electronic band structure, alterations in carrier transport properties, and the potential induction of novel quantum phenomena. Consequently, our results establish a strong foundation for future investigations aimed at harnessing and exploiting the EPC to control the bandgap of the semiconductors in other applicable systems, such as 2D quantum material systems.

**Materials and Methods**

**Material**

In our experiment, we employed a x-cut *BaF₂* crystal with a [100] orientation, which was obtained from a commercial company. The crystal had dimensions of 5 mm x 5 mm and was optically polished on both sides. The measured thickness of the crystal was approximately 200 μm. *BaF₂* crystallizes in the cubic $F\bar{m}3m$ space group and possesses a fluorite structure [39]. In this structure, $Ba^{2+}$ ions are bonded to eight equivalent $F^{1-}$ ions in a body-centered cubic arrangement, with all Ba-F bond lengths measuring 6.21 Å. The $F^-$ ions, in turn, are bonded to four equivalent $Ba^{2+}$ ions, forming a combination of corner-sharing and edge-sharing $FBa_4$ tetrahedra. Based on group theory analysis, the phonon modes in the primitive cell of a fluorite structure consist of three non-equivalent atoms, resulting in a total of nine phonon modes in the dispersion relations. Among these modes, three are acoustic modes. At the Γ point (the Brillouin zone center), there are three distinct optic phonon modes. These modes include a doubly degenerate infrared-active TO mode, a $T_{1u}$ mode, and a triply degenerate Raman-active mode $T_{2g}$. The $T_{2g}$ mode, which was the focus of observation and analysis in our study, represents the observed phonon mode.

**Optical characterization for the phonon mode**

In order to accurately assign the observed phonon mode, we conducted a Raman spectrum measurement on a single-crystal sample of *BaF₂*. This additional experimental technique allowed us to compare the vibration frequencies obtained from the observed coherent phonon mode, the results of density functional theory (DFT) calculations, and the Raman spectrum. For the Raman spectrum measurement, we utilized a custom-built system equipped with three ultra-narrow linewidth continuous wave lasers. These lasers were centered at wavelengths of 455 nm, 532 nm, and 780 nm, respectively. The system was designed to provide precise and controlled excitation for the Raman scattering process. To ensure a high signal-to-noise ratio in the recorded spectrum, we performed measurements using different pump lasers. In the main text, the presented spectrum corresponds to the measurement conducted with the 532 nm pulse. This specific configuration was selected based on considerations such as the experimental setup, laser characteristics, and desired spectral resolution.

**Experimental Setup and tr-HHG Spectroscopy Technique**

The experimental setup used in this study is depicted in Fig. 1a of the main text. It involved the utilization of two non-collinear laser pulses for the pump-probe HHG spectral detection geometry. The polarization of the probe pulse was fixed as P-polarized, while the polarization of the pump pulse could be varied by a halfwave plate. The crystal sample was mounted on a high precision rotational stage to ensure accurate alignment of the crystal orientations with respect to the incident laser pulses. This rotational stage allowed for precise control and adjustment of the crystal's orientation, enabling optimal interaction between the laser pulses and the crystal lattice. By aligning the crystal orientations, the experimental conditions could be carefully controlled and optimized for the desired measurements and observations. The pump laser was generated by a high-power Ti: Sapphire near-infrared (NIR) laser with a carrier wavelength of 800 nm. It had a total energy of 10 mJ and operated at a repetition rate of 10 kHz. The probe pulse, with a wavelength of 400 nm, was produced by frequency-doubling the 800 nm pulse. The pulse durations of the pump and probe were determined using a homemade transient grating frequency-resolved optical gating (TG-FROG) setup, yielding durations of approximately 22 fs and 25 fs, respectively. To focus the light pulses, a 25 cm focal length lens was employed, resulting in a beam size of around 25-40 μm, depending on the opening size of the iris. The HHG spectra were recorded using an imaging spectrometer equipped with a flat-field variable groove density grating and a CCD camera-coupled micro-channel plates (MCP) detector. The spectrometer covered a spectral range of 6 eV to 35 eV, limited by the collection angle of the extreme ultraviolet (EUV) spectrometer. The resolution of the spectrometer was approximately 0.05 eV, and it should be noted that the spectrum was not corrected for the grating's sensitivity. The non-collinear angle between the pump and probe pulses was estimated to be less than 3° based on the angles of the harmonics on the MCP detector. For time delay scans, a linearly closed-loop piezo stage was employed to precisely detune one of the pump-probe arms. The electric-field amplitudes of the pump and probe pulses inside the sample were estimated to be approximately $\sim 10^{12}$ W/cm² each. These estimations were derived using Fresnel's formula for S and P polarization under normal incident conditions, in conjunction with power-camera measurements.

**Data analysis**

The HHG and tr-HHG spectra presented in Fig. 1 and Fig. 2 were obtained by integrating the spectrum for a duration of 300 ms. To achieve a high spectral resolution, several factors were taken into consideration. Firstly, due to the 10 kHz repetition rate of the laser, the integration time of the camera was set to 500 ms. This ensured that each recorded spectrum was averaged over a significant number of laser shots, with an average of 200 shots per integration. This averaging process was performed at least 1,000,000 times for each time delay, resulting in a large dataset for analysis. The spectral resolution, defined by the Rayleigh criterion, was approximately 0.05 eV in all of our measurements. However, when determining energy shifts or relative energy differences, it is important to consider the standard error or standard error of the mean. In this case, the square root of the number of laser shots (1,000,000) or ~1000 should be used in the denominator to calculate the precision of the measurement. Furthermore, for fitting procedures that involve the full range of delays (at least 200 delays), which can be calculated by dividing it by another ~14 times (square root of 200). This takes into account the additional variability introduced by fitting over a larger range of time delays. Based on this estimation, it could be claimed that there is an enhancement of approximately ~1000 times in spectral resolution for determining energy shifts and ~14,000 times (1000×14) in spectral resolution for determining the effective energy shift. However, it is important to note that this estimation assumes no other systematic or random errors are present. To account for potential errors that may not have been considered, a conservative energy resolution of 0.001 eV was reported as the minimum spectral resolution in the determination of energy shifts. By employing these measures, the experimental setup aimed to achieve a high level of precision and accuracy in the determination of energy shifts and the effective bandgap, taking into account both the spectral resolution and potential sources of error.

**Density Functional Theory**

In this study, *ab* initio density functional theory (DFT) calculations were employed to determine the electron band structure, density of states (DOS), electron density of the *BaF₂* crystal. The band structure and DOS were computed using the Quantum Atomistix ToolKit (ATK) Q-2019.1252 commercial platform [41], which is based on first principles methods. The geometry optimization of the crystal was performed under the force field approximation, ensuring that the atomic positions were optimized to achieve a stable and energetically favorable configuration. Subsequently, the electronic structure calculations were conducted within the meta-generalized gradient approximation (MGGA) using the Perdew-Burke-Ernzerhof (PBE) parametrization. These approximations and parametrizations are widely used in DFT calculations and have been successful in describing the electronic properties of a variety of materials. To perform the calculations, the TB09 functionals and the Pseudo-Dojo pseudopotential were employed. The TB09 functionals are a set of exchange-correlation functionals that provide accurate descriptions of a wide range of materials. The Pseudo-Dojo pseudopotential is a type of pseudopotential that approximates the behavior of the core electrons in the crystal, allowing for more efficient calculations by reducing the computational complexity. Based on these calculations, the direct band gap at the Γ point of the *BaF₂* crystal was determined to be 10.2 eV, which closely matches the experimentally measured value of 10.5 eV [42]. This agreement between theory and experiment validates the accuracy of the DFT calculations and demonstrates the capability of the chosen approximations and parametrizations to describe the electronic properties of *BaF₂*. Supplementary Fig. 2a and 2b present the computed band structure and DOS, providing valuable insights into the electronic structure of the crystal.

**Density Functional Perturbation Theory**

Phonon quantities, such as phonon dispersion curves, vibrational modes were computed using density functional perturbation theory (DFPT) and the frozen phonon method of lattice displacements. These calculations were performed using the ATK platform. The lattice constants were obtained based on a zero-pressure optimized structure obtained using the Limited-memory Broyden–Fletcher–Goldfarb–Shanno (LBFGS) optimizer method. The LBFGS optimizer is an algorithm commonly used to minimize the total energy of a system and achieve stable atomic configurations. The calculations were performed using the PBE variant of the Generalized Gradient Approximation (GGA) exchange-correlation functional. The PBE functional is widely used in DFT calculations and provides reasonable accuracy for a range of materials. A cut-off energy of 750 eV was used to obtain converged calculation results. This cut-off energy determines the maximum energy for plane waves included in the calculations and affects the accuracy of the computations. A higher cut-off energy allows for a more complete representation of the wavefunctions, but also requires more computational resources. The electronic BZ integrals were computed using a 4×4×4 Monkhorst-Pack mesh. The Monkhorst-Pack method is a technique used to sample the Brillouin zone of a crystal lattice in electronic structure calculations. A 4×4×4 mesh means that the Brillouin zone is divided into a grid of points in reciprocal space, and the electronic integrals are evaluated at these points. This choice of mesh size provided convergence in the calculations, as indicated by a force constant of less than 1 meV/Å.

**Dynamical Quantum Model for HHG radiation incorporating phonons**

The band gap of *BaF$_2$* is approximately 10 eV, and it exhibits strong excitonic features in the vicinity of the band gap [42]. These excitonic features play a dominant role in the optical response within this frequency range. In contrast, the central frequency of the probe pulse used in the study (3.1 eV) is significantly higher than the resonant band gap transition and well above the THz frequency range, where the polarization and current sources of high-harmonic generation (HHG) are comparable. Based on these considerations, the study concludes that the optical signals observed are primarily governed by excitonic transitions rather than quasiparticle band-to-band transitions, as suggested by previous published results. This distinction is important in understanding the underlying physical processes and interpreting the experimental observations. As a result, the equations of motion governing the dynamics of the system can be simplified, taking into account the dominant excitonic transitions. This simplification allows for a more focused and accurate analysis of the experimental data and provides insights into the behavior of excitons in the *BaF$_2$* crystal system [37, 43-44].

$$\frac{d}{dt}f_0 = -i\varepsilon(t)(p - p^*), \quad (4)$$

$$\frac{d}{dt}p = -i\Omega(t)p + i\varepsilon(t)(1-2f_0), \tag{5}$$

$$\Omega(t) = \frac{1}{\hbar}(E_{1s} + M_p(t)), \tag{6}$$

$$M_p(t) = \sum_i V_i \cos(\omega_{i,0}t + \varphi_i)\exp\left[-\frac{t}{\tau_i}\right] \tag{7}$$

with coefficients $V_i$ represent the *EPC* strength and we have added a phenomenological damping time $\tau_i$ of the respective phonon mode. To obtain the optical signals, we solve the equation of motion for the polarization induced by the probe pulse (Eq. (4-6)), taking into account a phenomenological dephasing time $T_2$. Subsequently, we calculate the Fourier transform of the microscopic polarization p, which serves as the source of the emitted radiation, i.e., E(ω)∝p(ω).

In the extension from a two-level quantum model to a two-band model, the semiconductor Bloch equations (SBEs) are considered. Under a homogeneous excitation, and taking into account the phonon mode that is only excited at **Q** = 0 (where **Q** represents the phonon wavevector), the SBEs can be written as follows [45]:

$$i\hbar\frac{\partial}{\partial t}p_\mathbf{k} = \left(\epsilon_\mathbf{k}^c - \epsilon_\mathbf{k}^v + M_p(t) - i\frac{\hbar}{T_2}\right)p_\mathbf{k} - \left(1 - n_\mathbf{k}^e - n_\mathbf{k}^h\right)\mathbf{d}_\mathbf{k}\cdot\mathbf{E}(t) + ie\mathbf{E}(t)\cdot\nabla_\mathbf{k}p_\mathbf{k}, \tag{8}$$

$$\hbar\frac{\partial}{\partial t}n_\mathbf{k}^{e(h)} = -2\,\mathrm{Im}\left[\mathbf{d}_\mathbf{k}\cdot\mathbf{E}(t)p_\mathbf{k}^*\right] + e\mathbf{E}(t)\cdot\nabla_\mathbf{k}n_\mathbf{k}^{e(h)}, \tag{9}$$

Where $P_k$ and $n_k$ represent the interband coherent and electrons (hole) populations is the energy of the electrons (holes) in the valence (conduction) band, $T_2$ is the dephasing time of interband polarization, $\mathbf{d_k}$ is the dipole matrix element corresponding to the transition between two bands, and $\mathbf{E}(t)$ is the electric field of pump or probe laser pulses.

The macroscopic time-dependent interband polarization **P**(t) and intraband current density **J**(t) are written as

$$\mathbf{P}(t) = \sum_\mathbf{k}\left[\mathbf{d}_\mathbf{k}p_\mathbf{k} + \text{c.c.}\right], \tag{10}$$

$$\mathbf{J}(t) = \sum_{v(c),\mathbf{k}} e v_{\mathbf{k}}^{v(c)} n_{\mathbf{k}}^{v(c)}, \tag{11}$$

where $v_{\mathbf{k}}^{v(c)}$ is the group velocity of the valence (conduction) band, which is defined through $v_{\mathbf{k}}^{v(c)} = \nabla_{\mathbf{k}} \epsilon_{\mathbf{k}}^{v(c)} / \hbar$. Finally, the total spectrum of emitted radiation becomes

$$S(\omega) \propto |\omega \mathbf{P}(\omega) + i \mathbf{J}(\omega)|^2. \tag{12}$$

Where the $\mathbf{P}(\omega)$ and $\mathbf{J}(\omega)$ are the FFT of the interband polarization $\mathbf{P}(t)$ and intraband currents $\mathbf{J}(t)$, the dipole moment matrix elements are calculated by $\mathbf{k} \cdot \mathbf{p}$ perturbation theory, which is

$$\mathbf{d}^{\lambda'\lambda}(\mathbf{k}',\mathbf{k}) = \mathbf{d}^{\lambda'\lambda}(0,0) \delta_{\mathbf{k}\mathbf{k}'} \frac{\epsilon_0^\lambda - \epsilon_0^{\lambda'}}{\epsilon_k^\lambda - \epsilon_{k'}^{\lambda'}}. \tag{13}$$

**Acknowledgments:** It is our pleasure to acknowledge fruitful discussions with Daniel Wigger, Doris Reiter, Tilmann Kuhn. We gratefully acknowledge funding support by the Department of Physics, Faculty of Science, HKU, the RGC ECS project 27300820, the GRF project 17315722, and the Area of Excellence project AoE/P-701/20.

**Author contributions**: Jicai Zhang, Ziwen Wang, Tien Dat Tran, Wenhao Yu, Chong Zhang, Wenqi Xu and Marcus Lo contributed to building the experimental set-up; Tien Dat Tran and Marcus Lo contributed the software of data acquisition. Jicai Zhang conducted the experiments and performed the data analysis and simulations. Jicai Zhang and Tran Trung Luu write the manuscript; Tran Trung Luu supervised the project. All authors discussed and interpreted the experimental data.

**Competing interests:** All authors declare that they have no competing interests.

**Data and materials availability:** All data are available in the main text or the supplementary materials.

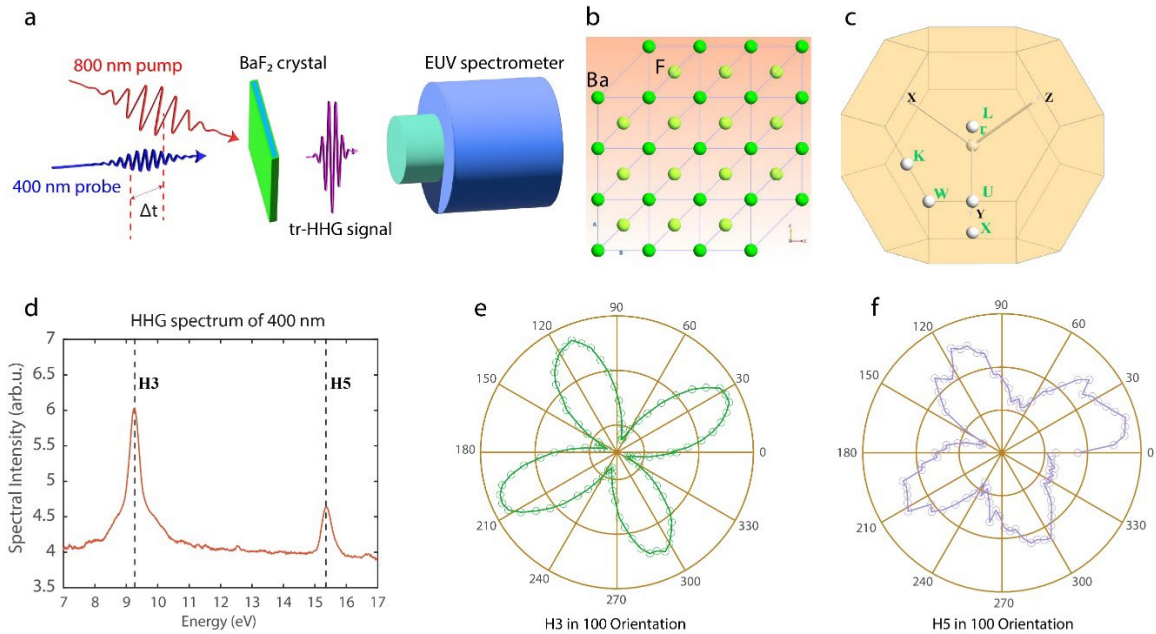

**Fig. 1 | Experimental apparatus and HHG spectra measurement. a** Schematic of the experimental setup: two ultrashort laser pulses with centre wavelengths at 800 nm (pump ~30 fs) and 400 nm (probe ~25 fs) are used to form a non-collinear pump-probe HHG detection geometry. The HHG spectra from the probe are recorded by an EUV spectrometer placed downstream of the sample. **b** Super cell of $BaF_2$, which has a face-centered cubic (*fcc*) lattice structure, with each $Ba^{2+}$ ion surrounded by eight $F^-$ ions and each $F^-$ ion surrounded by four $Ba^{2+}$ ions. **c** Brillion zone of the $BaF_2$. **d** Static HHG spectrum of the 400nm, which includes 3$^{rd}$ and 5$^{th}$ low order harmonics, the vertical dashed line indicates the central energy of the two harmonics. **e** and **f** Crystal orientation dependent 3$^{rd}$ and 5$^{th}$ harmonic spectrum (H3, H5) of the $BaF_2$ in the [100] cut direction, respectively.

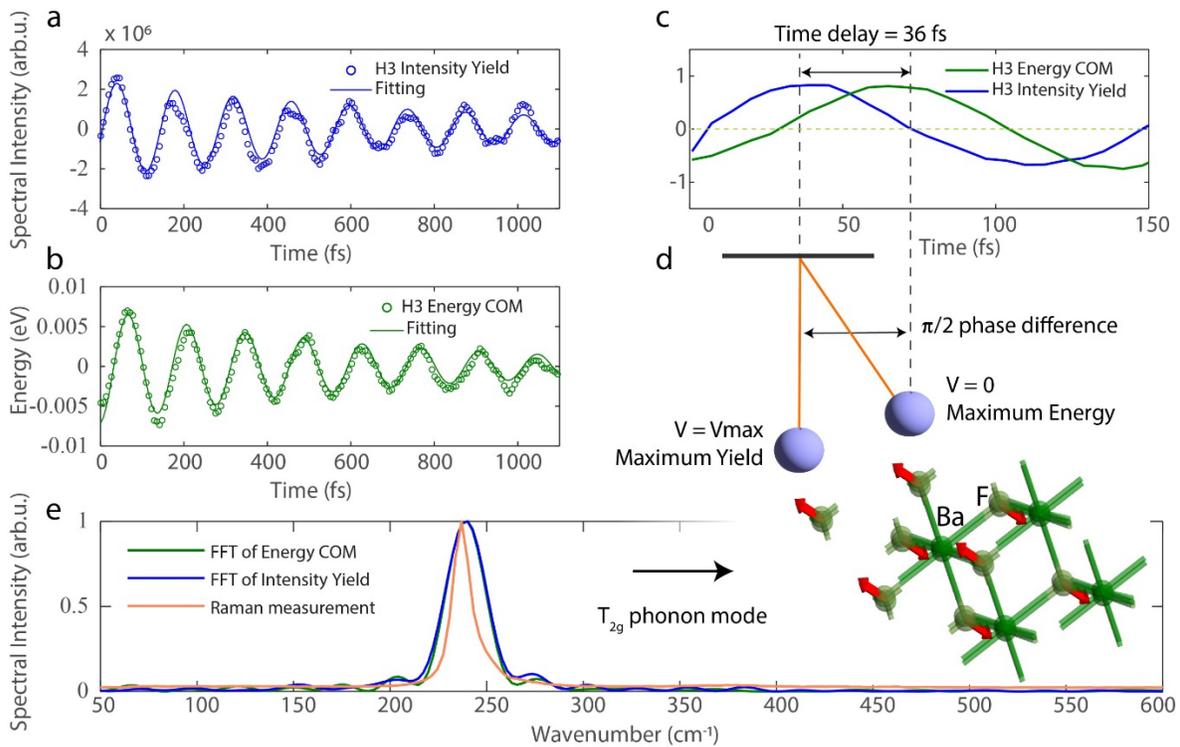

**Fig. 2 | Coherent phonon oscillation dynamics probed with tr-HHG spectroscopy. a** Time-resolved intensity yield of H3 (blue circles) and its fitting (blue curve). **b** Center of mass of H3 photon energy (dark green circles) and its fitting (dark green curve), subtracted by the average value. **c** Same as **a** and **b** but with a shorter time window. **d** The simple pendulum model indicates the initial phase difference, where the maximum and minimum velocities of the pendulum correspond to the maximum harmonic yield and maximum harmonic energy modulations, respectively. **e** Phonon spectra were measured using different techniques: FFT of the center COM (dark green curve), FFT of the intensity yield (blue curve), and Raman measurement (light orange curve). The intensities of all traces are normalized to their maximum.

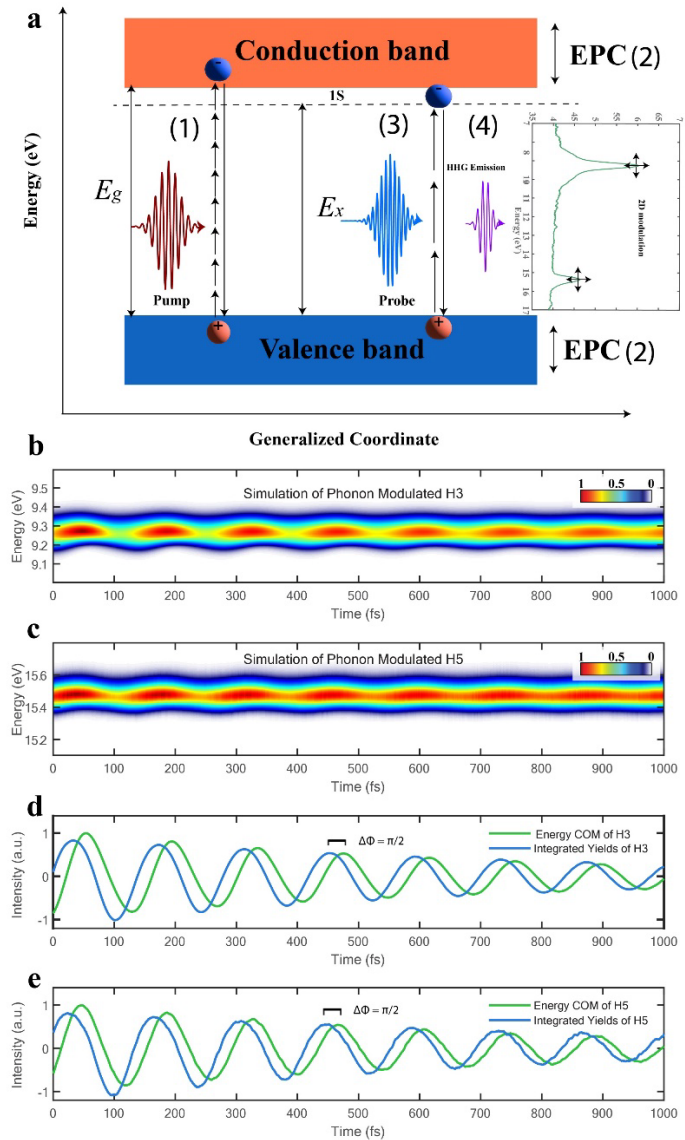

**Fig. 3 | Phonon perturbed physical picture and simulation of the phonon dynamics by two-level quantum model. a** Energy space excitation and probing of the coherent phonon dynamics, $E_g$ and $E_x$ represent the band gap energy and optical excitation exciton energy. **b** and **c** Simulated tr-HHG spectrum trace of 3rd and 5th harmonics based on the two-level quantum model, respectively. The colorbar is linearly scaled. **d** and **e** The time delayed 3rd and 5th harmonic as the integrated yields (blue color) variation and the integrated COM of energy (green color) variation, respectively.

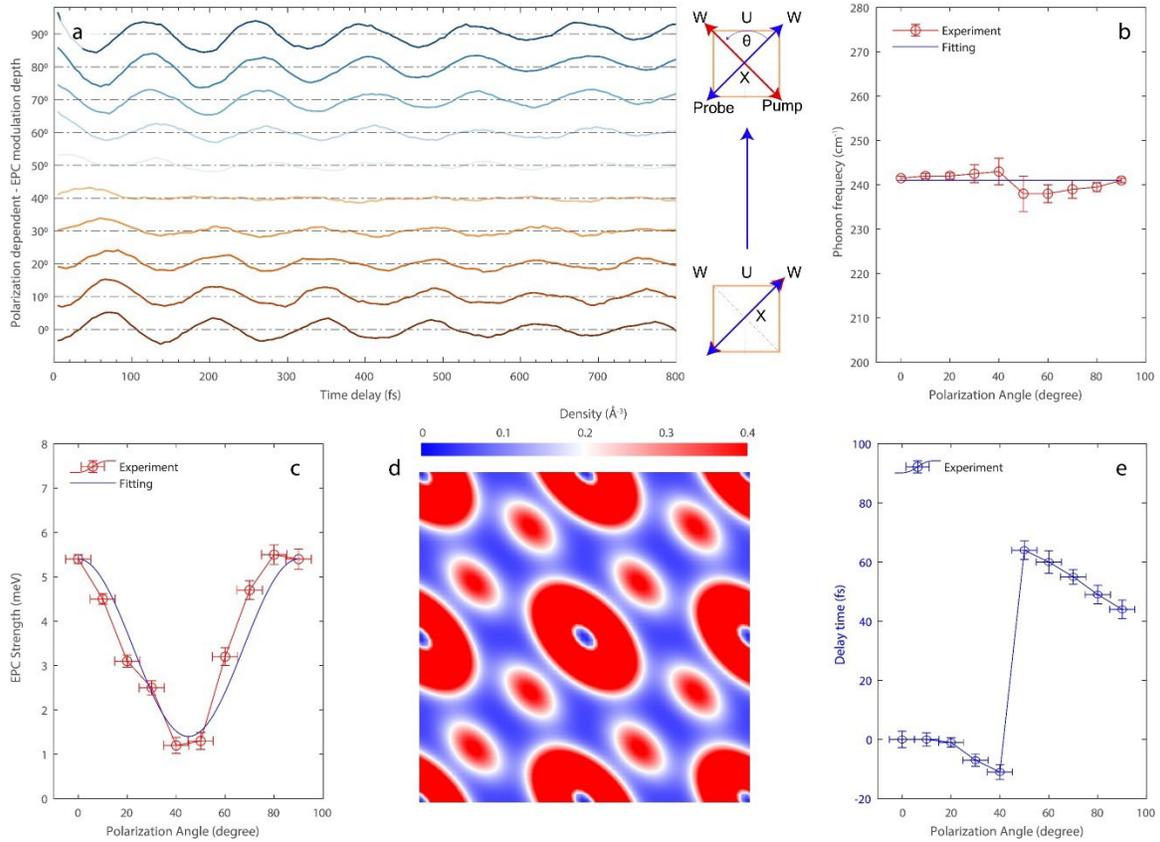

**Fig. 4 | Pump polarization dependent EPC dynamics variation. a** From bottom to top panel (red to blue curve) indicating the COM of tr-THG spectrum varying with the relative polarization angle (from 0 to 90°) between pump and probe pulses via rotating the half waveplate in the pump beam. The right-side drawing shows the pulse polarization direction regarding the crystal orientation (high symmetry X-W and X-U path) in BZ. The probe pulse is always aligned to X-W orientation. **b** Experimentally extracted oscillation frequency under different pump pulse polarizations (red circles) and its linear fit (blue line). **c** Extracted EPC strength at different polarization angles (red circles) and its cosine fit (blue curve). **d** The two-dimensional electron density distribution in the x-cut plane on barium atoms obtained from the DFT calculation, corresponds to valence electrons occupying the orbitals below the Fermi level (E = 0 eV). **e**, Extracted relative delay time based on the fitted initial phase of phonon oscillations under different polarizations. The error bar accounts for both statistical and systematic uncertainties.

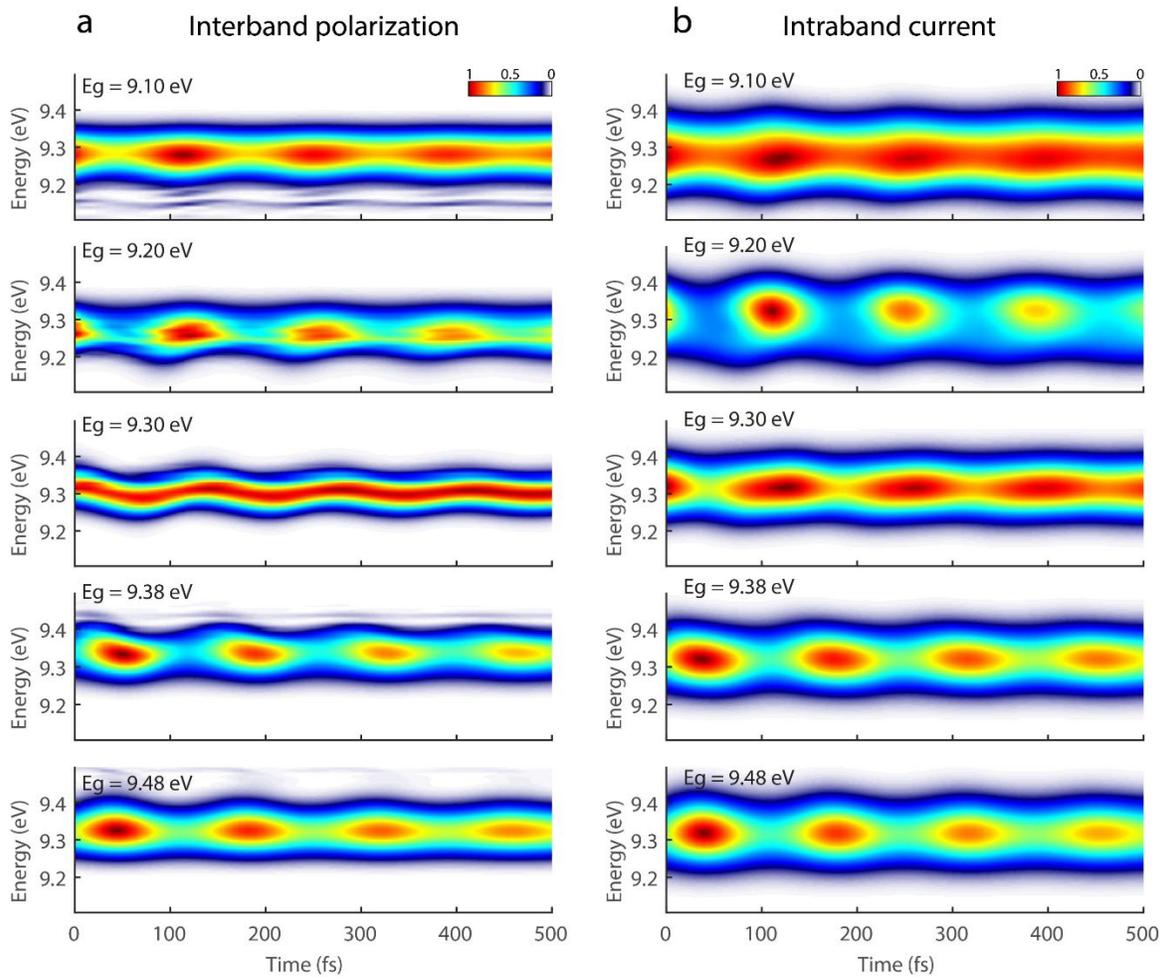

**Fig. 5 | Simulated time-delayed interband and intraband THG spectra variation under different bandgap values. a** Time-resolved interband THG spectrum with increasing the bandgap energy from 9.0 eV to 9.48 eV. **b** Same as **a** but showing only intraband current contributions.

Supporting Information for

# All-Optical Manipulation of Band Gap Dynamics via Electron-Phonon Coupling

Jicai Zhang[†], Tien-Dat Tran, Ziwen Wang, Wenhao Yu, Chong Zhang, Marcus Lo, Wenqi Xu & Tran Trung Luu[†].

[1]Department of Physics, The University of Hong Kong; Pok Fu Lam Rd, Hong Kong SAR, China.

[†]Corresponding author: jczhang@hku.hk; ttluu@hku.hk

**This PDF file includes:**

Figures S1 to S6

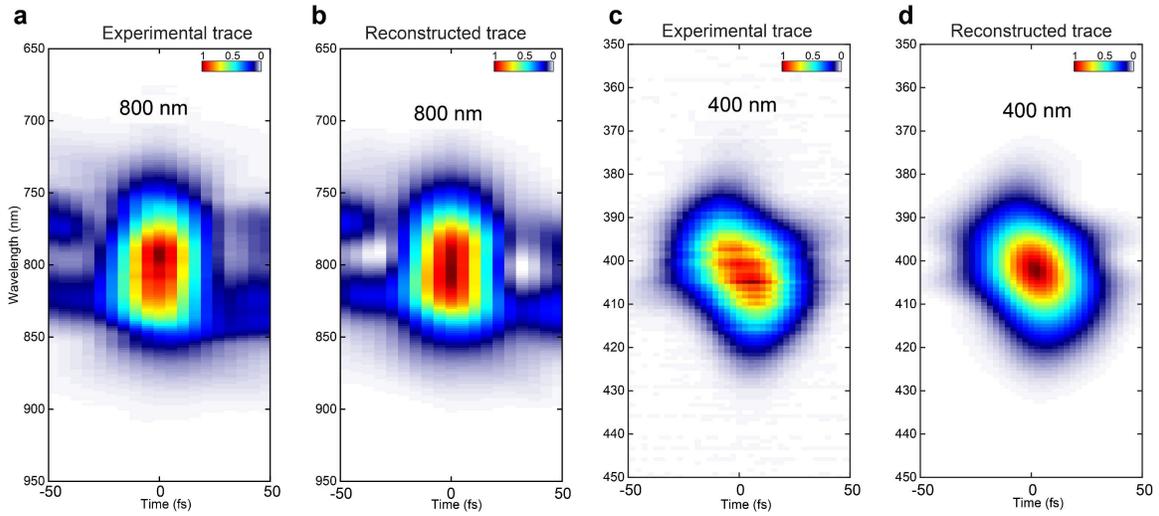

**Supplementary Figure 1 | Measured and reconstructed TG-FROG Spectrum Traces of 800 nm pump pulse.** (**a** and **b**) and (**c** and **d**) The measured and reconstructed pulse duration of the pump (800 nm) and probe (400 nm) pulses, respectively.

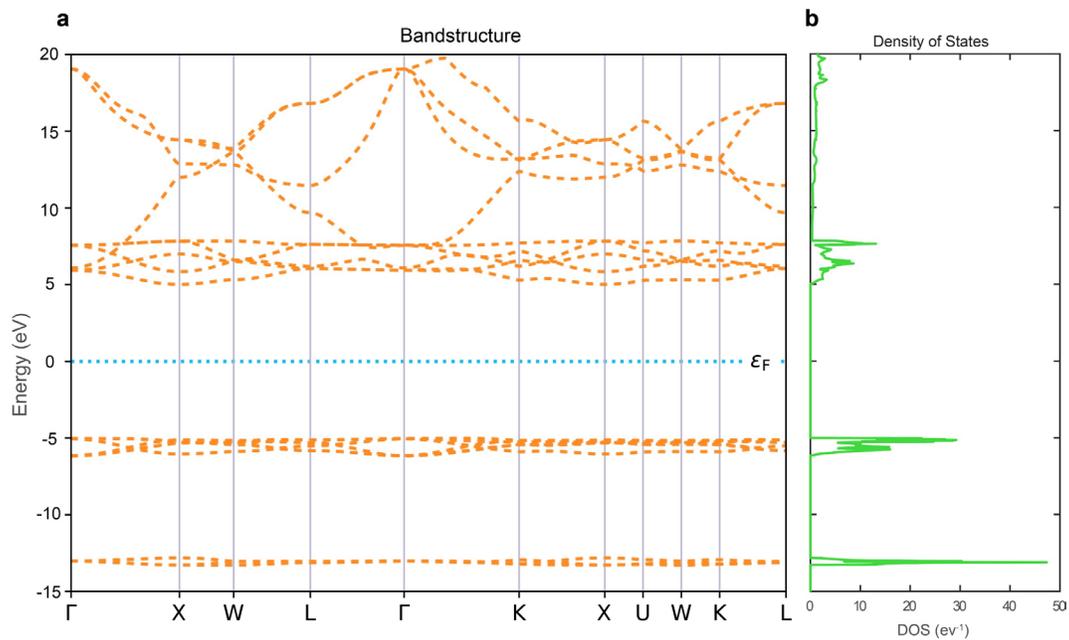

**Supplementary Figure 2 | Electronic Band structure of BaF$_2$ crystal along high symmetrical paths.** (**a** and **b**) Calculated electronic band structure and density of states from DFT.

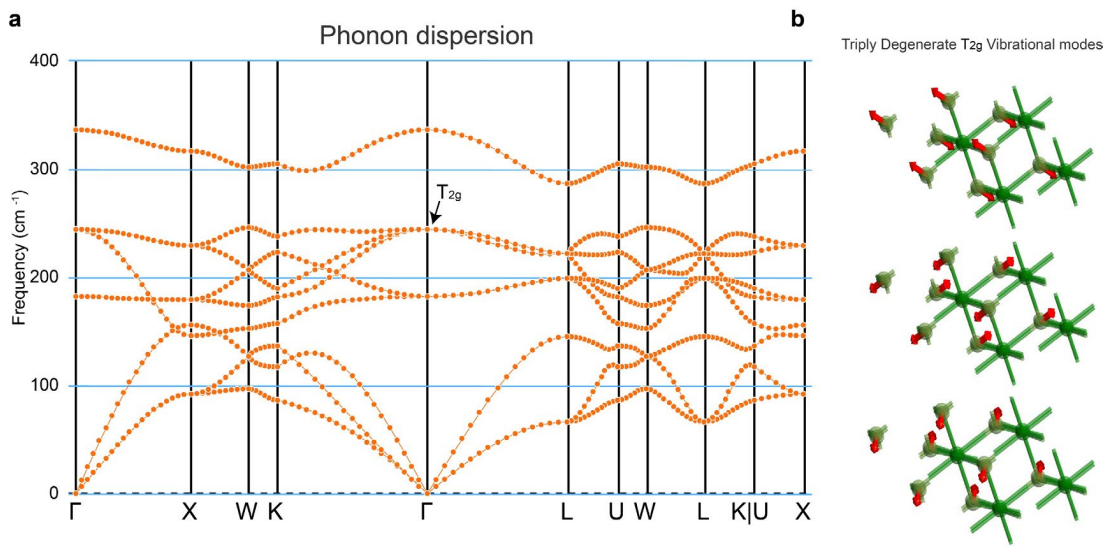

**Supplementary Figure 2 | Phonon dispersion curve of BaF$_2$ crystal**. **a** Phonon dispersion that were calculated from DFPT. **b** Triply degenerate vibrational mode of T$_{2g}$ at the Γ and L points.

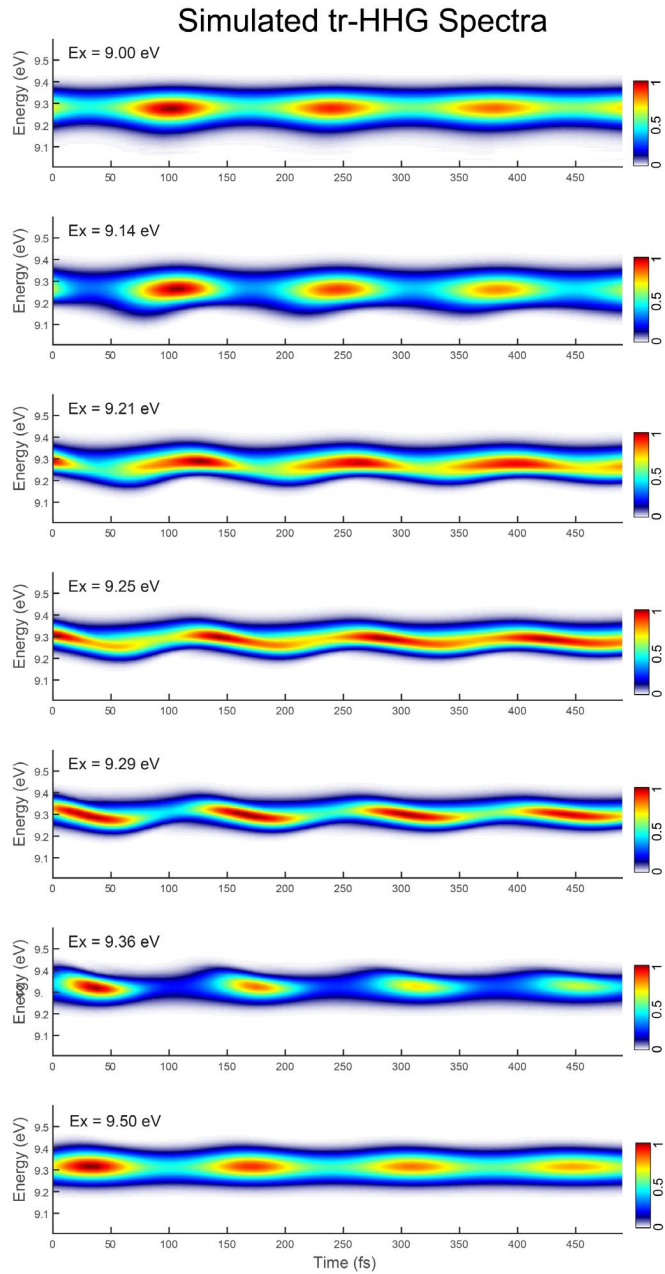

**Supplementary Figure 4 | Simulated exciting exciton energy $E_x$ dependent tr-THG spectrum trace.** Simulated tr-THG spectrum trace by varying the exciting $E_x$ energy from 9.0 eV to 9.5 eV (from the top to the bottom). The colorbar is plotted as the linear scale.

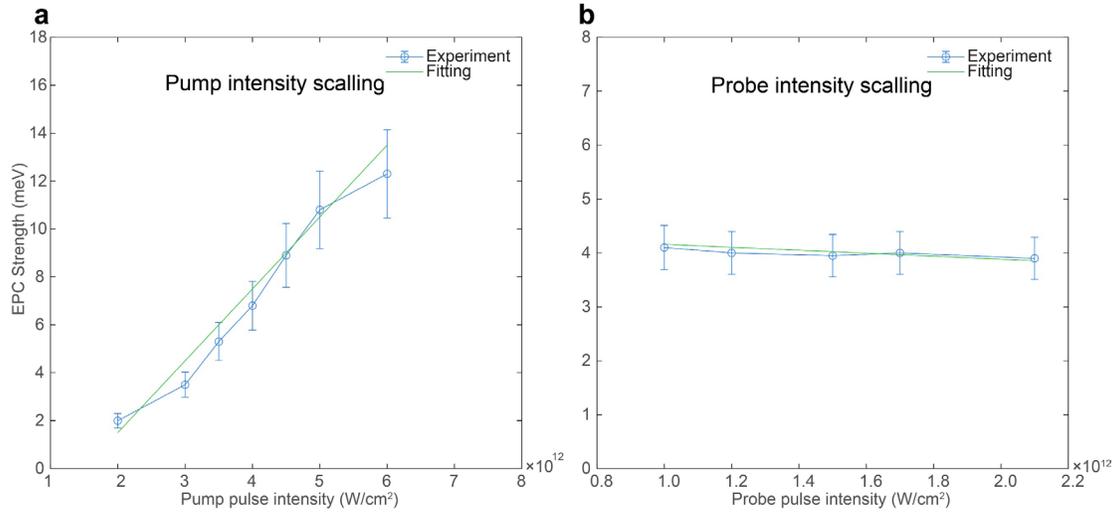

**Supplementary Figure 5 | Intensity scaling of ECP strength under different pulse strengths**. **a** and **b** Recorded ECP strength using different pump and probe intensities when fixing the probe and pump pulses at 2×12 W/cm$^2$ and 2×10$^{12}$ W/cm$^2$, respectively. The green solid lines represent the linear fitting.

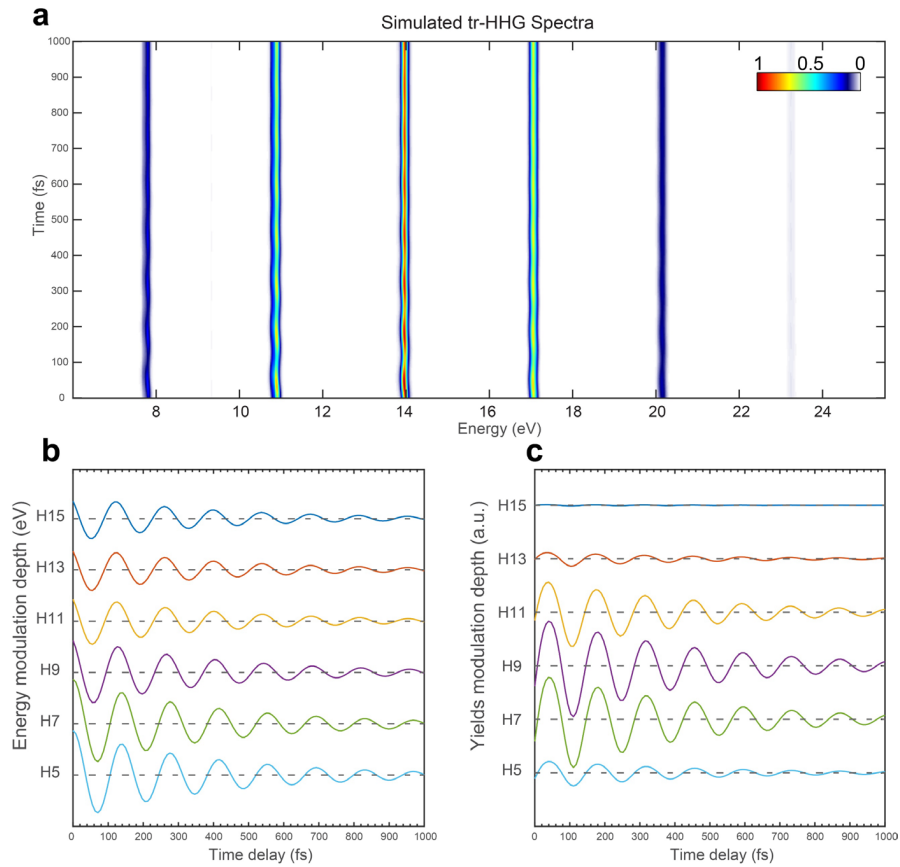

**Supplementary Figure 6 | Simulated tr-HHG spectra from a two-band model pumped by an 800 nm wavelength pulse. a** Simulated tr-HHG spectrum from the 5[th] to the 15[th] order of harmonics. **b** and **c** The integrated yields and the COM spectrum of the corresponding trace **a**, respectively.